\documentclass[prl,aps,twocolumn,showpacs]{revtex4-1}

\pdfoutput=1 
\usepackage{graphics} 
\usepackage{graphicx} 

\begin{document}

\title{Glass-forming ability of Lennard-Jones trimers}

\author{Ulf R. Pedersen} 
\email{urp@ruc.dk} 
\affiliation{Glass and Time, IMFUFA, Department of Science and Environment, Roskilde University, Postbox 260, DK-4000 Roskilde, Denmark}

\begin{abstract}
Melting temperatures at ambient pressure of systems of isosceles Lennard-Jones trimers with angles ranging from 70 degrees to 100 degrees are determined. Two crystal structures are considered: a distorted body centered cubic structure and a distorted face centered cubic structure with preferred angles of 77 and 96 degrees, respectively. Liquid dynamics are slowed down when the angle is increased. A trimer angle of 83 degrees yields the largest distance between isochrones and the melting temperature, 
suggesting that this value gives the optimal glass-forming 
ability. It is conjectured that 
better glass-formers may be found at angles larger than the ones considered in this study.
\end{abstract}

\maketitle

In the last decades, the use of computer simulations played an increasingly important role in investing underlying assumptions and predictions of theories.
Well-studied simulation models, such as the Kob-Andersen binary LJ mixture \cite{kob1994}, the Wahnstr{\"o}m LJ mixture \cite{wahnstrom1991} or the Lewis-Wahnstr{\"o}m isosceles LJ trimers \cite{lewis1993,lewis1994} (one of the trimers studied in this paper), are characterized by being computationally cheap and not prone to crystallization in comparison to the single component LJ model. With increasing computer power, however, simulations nowadays reach timescales where these models crystallize \cite{pedersen_2010wablj,toxvaerd2009,pedersen2011_lwotpCry}. Thus, there is a need for models that inherit the simplicity of the well-studied models, while being better glass-formers. In this paper we investigate crystalline stability of isosceles LJ trimers similar to the model suggested by Lewis \& Wahnstr{\"o}m \cite{lewis1993,lewis1994}. The strategy is to design a good glass-former by changing the trimer angle. For this, the crystal stability is identified by computing the melting temperature for various angles (
at ambient pressure considering two crystal structures). We note that Molinero et. al \cite{molinero2006} designed a good glass-former by changing a parameter controlling the tetrahedral characteristics of a mono-atomic silicon model.

For the originally 
suggested angle of $75^\circ$, trimers 
crystallize into a structure where LJ particles occupies a near body centered (BC) cubic lattice structure (consistent with the finding of References \cite{pedersen2011_lwotpCry,pedersen2011_lwotp}). At angles near $90^\circ$, however, near close-packed face centered (FC) cubic structures are more stable. At $83^\circ$ destabilization of the crystal is maximized. Moreover, liquid dynamics are slowed when the angle is increased.

\begin{figure} 
\begin{center} 
\includegraphics[width=1.0\columnwidth]{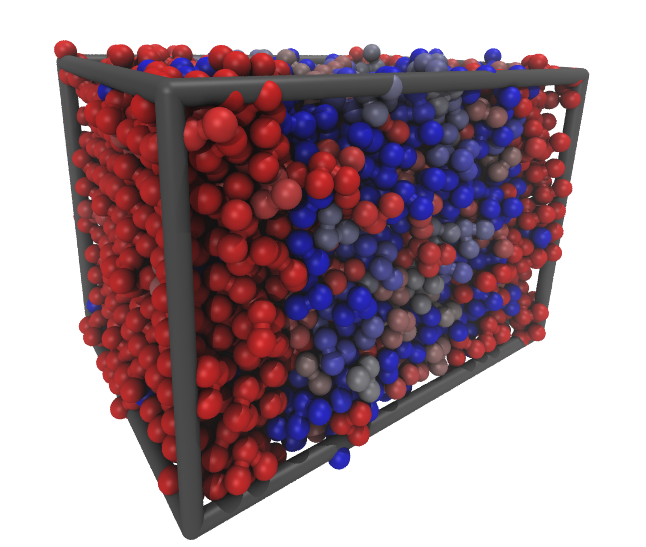}
\caption{\label{box3D} Configuration of LJ trimers ($\theta=75^\circ$) in a periodic orthorhombic box. A field biasing two-phase configurations have been added. The red molecules are in a crystalline inviroment, while the blue are in a liquid inviroment.}
\end{center} 
\end{figure}

\section{Crystallization of LJ trimers} 

\begin{figure} 
\begin{center} 
  \includegraphics[width=0.9\columnwidth]{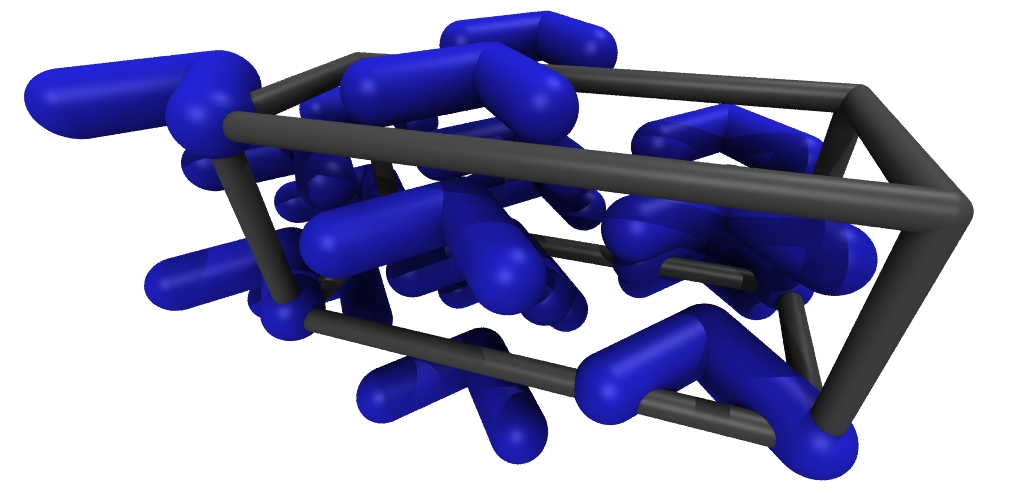} \\
  \includegraphics[width=0.7\columnwidth]{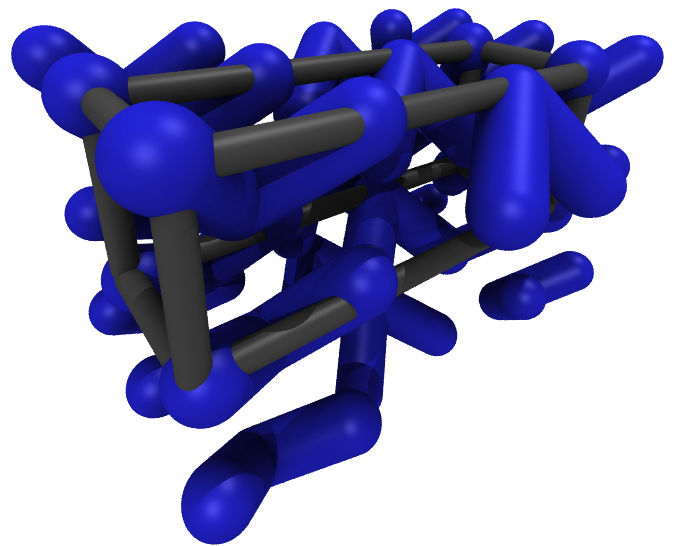}
\caption{\label{triuc} Close packed structures of isosceles LJ trimers. LJ particles (on triangle corners) occupy sites akin to the face centered cubic (upper) and body centered cubic (lower) structure (Table \ref{tbl_cells}). Gray boxes outline orthorhombic unit cells containing four and two trimers respectively. These structures may accommodate a range of trimer angles. }
\end{center}
\end{figure}

\begin{figure} 
\begin{center} 
  \includegraphics[width=1.0\columnwidth]{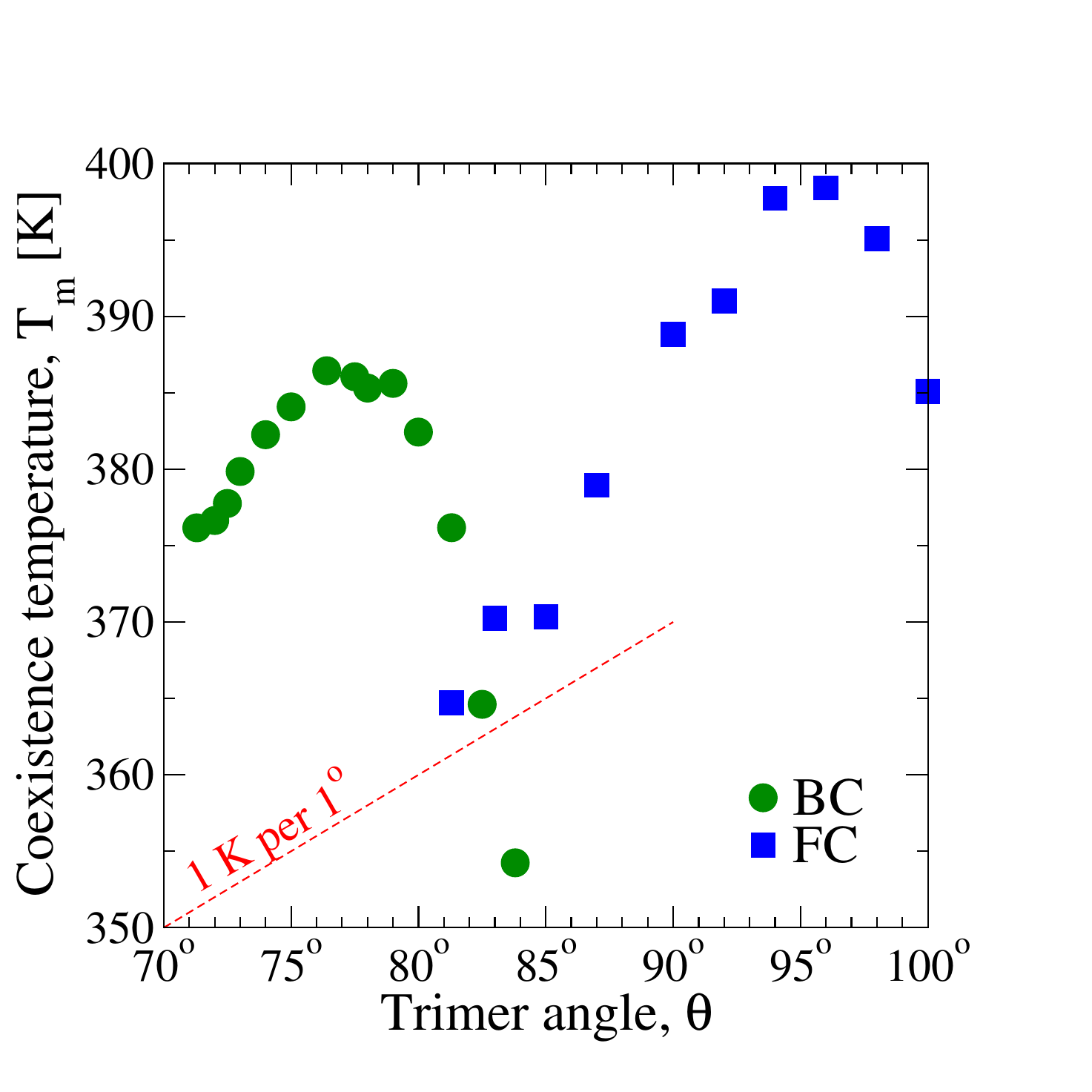}
\caption{\label{trimerTm} Solid-liquid coexistence temperature at ambient pressure as a function of trimer angle.}
\end{center} 
\end{figure} 

\begin{figure} 
\begin{center} 
  \includegraphics[width=1.0\columnwidth]{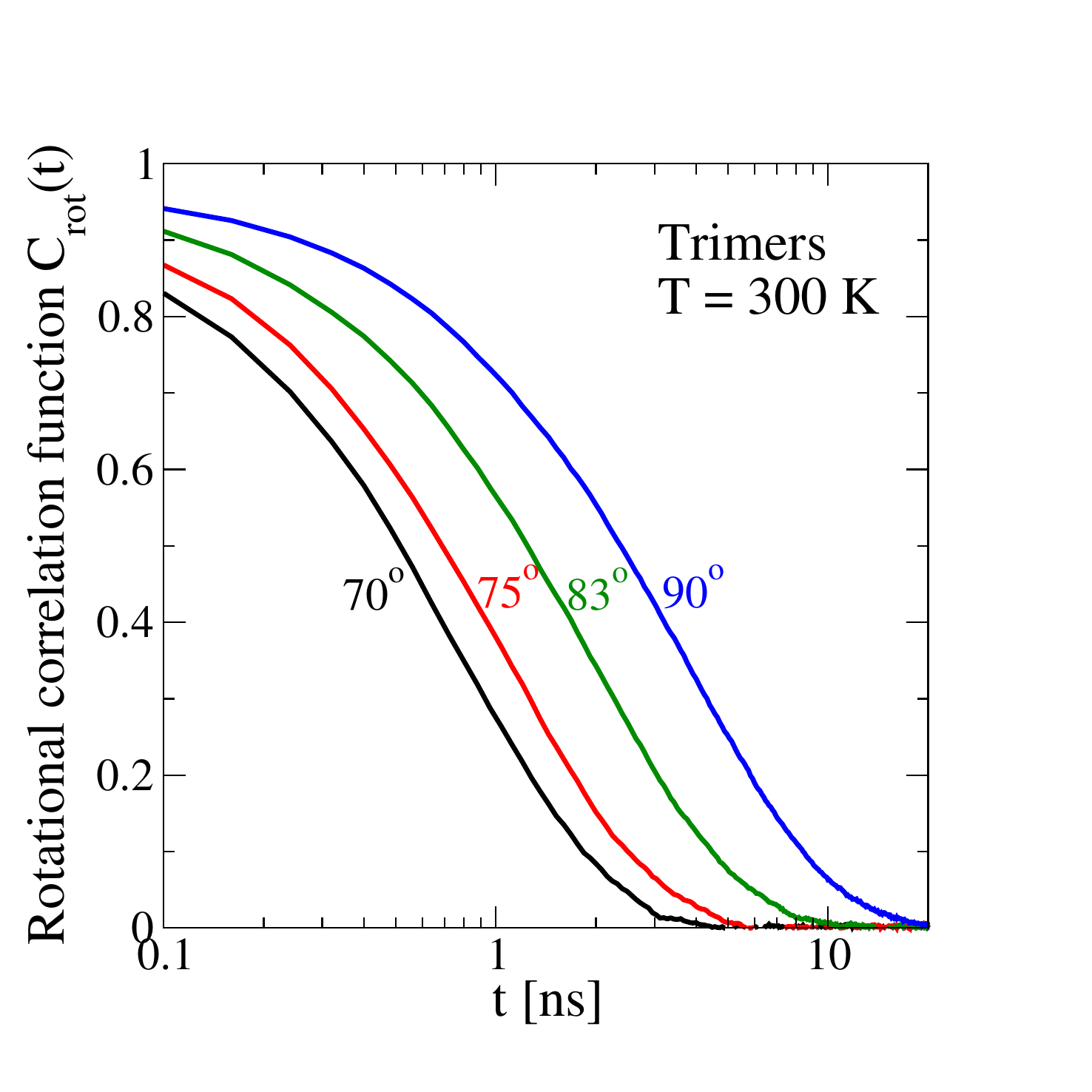}
\caption{\label{rotacf} Rotational autocorrelation function of the trimer long-bond of trimers with angles of $\theta=\{70^\circ,75^\circ,83^\circ,90^\circ\}$ at $T=300$ at ambient pressure. An increase of the angle slows dynamics.}
\end{center} 
\end{figure} 

\begin{figure} 
\begin{center} 
  \includegraphics[width=1.0\columnwidth]{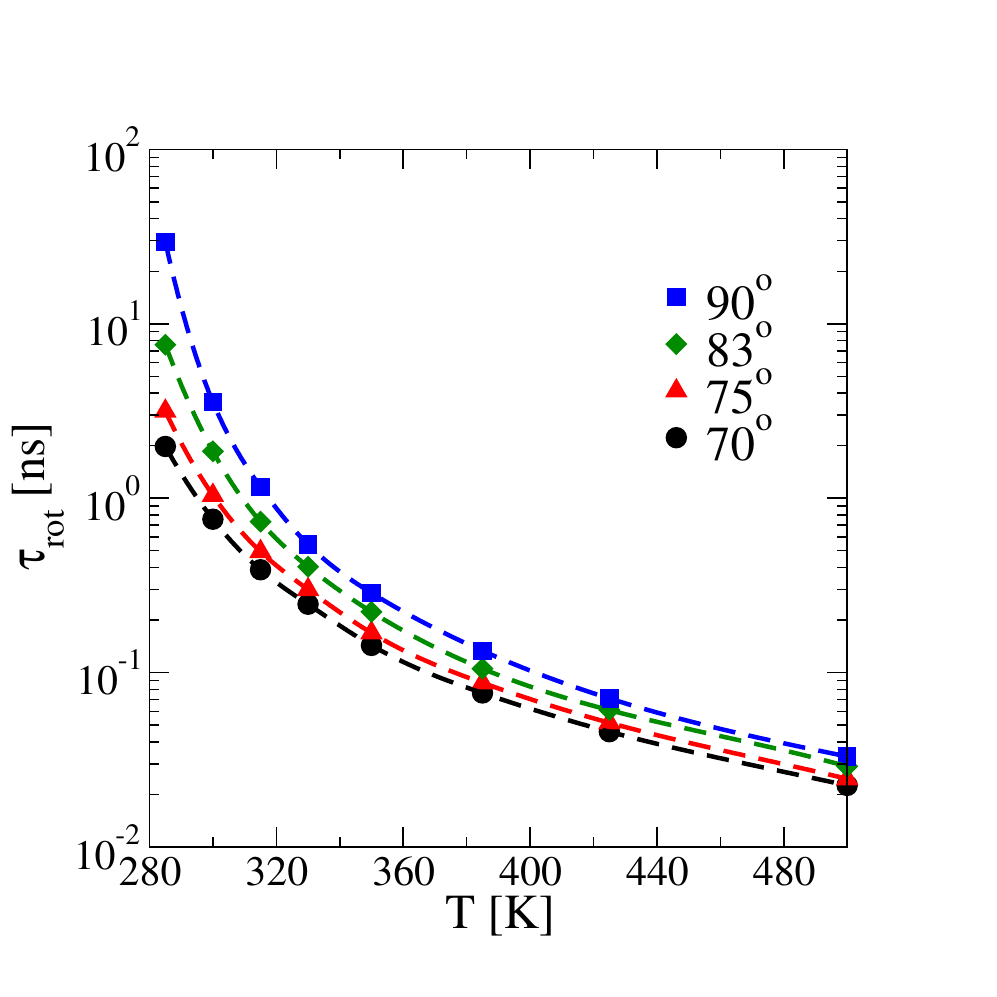}
\caption{\label{tau} Rotational correlation time as a function of temperature of trimers with angles $\theta=\{70^\circ,75^\circ,83^\circ,90^\circ\}$. The insert shows isochrones in the temperature-angle plane. }
\end{center} 
\end{figure} 

Next, we apply the interface pinning method to LJ trimers. Consider molecules of three LJ particles placed in the corners of a isosceles triangle with two sides of length $\sigma$ spanning an angle $\theta$. Bonds and angles are constrained using harmonic interactions. 

In 1993 Lewis \& Wahnstr{\"o}m \cite{lewis1993,lewis1994} suggested an angle of 75$^\circ$ as a model of a glass former, specifically ortho-terphenyl (OTP). Nowadays computers, however, reach timescales where the supercooled liquid of this model crystallizes into a structure where LJ particles occupies a near body centered (BC) cubic lattice \cite{pedersen2011_lwotpCry,pedersen2011_lwotp}. The optimal angle in the body centered cubic is $2\sin^{-1}(1/\sqrt{3})\simeq70.5^\circ$ \cite{pedersen2011_lwotpCry}, however, since the typical LJ distance is larger than $\sigma$ ($r_\textrm{min}/\sigma=2^\frac{1}{6}\simeq1.12$), the optimal angle larger. From an approximate analytic expression of the $p=0$, $T=0$ stability, it was argued that the Lewis-Wahnstr{\"o}m angle is surprisingly close to the optimal angle of 76.4$^\circ$. This was the direct motivation to address the question: ``What is the angle dependency of crystal stability relative to the liquid?''. We will limit ourself to $70^\circ<\theta<100^\circ$ and 
adapt LJ parameters suggested by Lewis \& Wahnstr{\"o}m: $\sigma=4.83$ nm, $\varepsilon=(600\textrm{ K})\times k_B$ and $m=76.768$ u.

\subsection{Stability of close packed structures}
First we need to construct candidates for close packed structures. We will use a philosophy of simplicity and intuition and leave a more systematic search for future investigations. When the trimer angle is near those between close packed spheres, $60^\circ$ and $90^\circ$, LJ particles may occupy sites akin to hexagonal close packed or face centered (FC) cubic. Triangles may decorate these lattices in numerous ways. Structures made up of lines of ``herring bones'' \cite{pedersen2011_lwotpCry} may accommodate a range of angles bond lengths and LJ-pair distances while keeping an orthorhombic unit cell. Figure \ref{triuc} shows two such structures and their orthorhombic unit cells. Super-cells was equilibrated at $p=0$, and varies $T$ and $\theta$. Average molecule coordinates in orthorhombic unit-cells are listed in Table \ref{tbl_cells}.

Figure \ref{trimerTm} shows the coexistence temperature for closed packed structures with respect to the liquid (at $p=0$) computed by interface pinning method \cite{pedersen2013,pedersen2013b}. The optimal angle for the BC structure is $\sim78^\circ$ in good agreement with the findings of Ref. \cite{pedersen2011_lwotpCry}. Similar, the optimal angles for the FC structure is $95^\circ$ rather than $90^\circ$. As for the BC structure, this widening of the angle is due intermolecular neighbor distances being larger than intramolecular distances.

\begin{table}
\caption{ \label{tbl_cells} Orthorhombic trimer unit cells.}
\ \\ 
BC: Pmn2$_1$ (31); $p=0$; $T=385$; $\theta=75^\circ$;\\$a=3.748$; $b=1.322$; $c=1.196)$;\\
\begin{tabular}{ l | c c c c c c }
          \#(atom)  & x/$a$ & y/$b$ & z/$c$ \\ 
\hline
1(base) & 0.000 & 0.000 & 0.000 \\ 
1(top)  & 0.163 & 0.450 & 0.438 \\ 
1(base) & 0.326 & 0.000 & 0.000 \\ 
2(base) & 0.500 & 0.500 & 0.500 \\ 
2(top)  & 0.663 & 0.050 & 0.938 \\ 
2(base) & 0.826 & 0.500 & 0.500 \\ 
\end{tabular}
\ \\ \ \\ \ \\
FC: Fmm2 (42); $p=0$; $T=385$; $\theta=83^\circ$; \\$a=4.096$; $1.637$; $c=1.759$;\\
\begin{tabular}{ l | c c c c c c }
          \#(atom)  & x/$a$ & y/$b$ & z/$c$ \\ 
\hline
1(base) & 0.000 & 0.000 & 0.000 \\ 
1(top)  & 0.162 & 0.457 & 0.000 \\ 
1(base) & 0.324 & 0.000 & 0.000 \\ 
2(base) & 0.500 & 0.500 & 0.000 \\ 
2(top)  & 0.662 & 0.957 & 0.000 \\ 
2(base) & 0.824 & 0.500 & 0.000 \\ 
3(base) & 0.000 & 0.500 & 0.500 \\ 
3(top)  & 0.162 & 0.957 & 0.500 \\ 
3(base) & 0.324 & 0.500 & 0.500 \\ 
4(base) & 0.500 & 0.000 & 0.500 \\ 
4(top)  & 0.662 & 0.457 & 0.500 \\ 
4(base) & 0.824 & 0.000 & 0.500 \\ 
\end{tabular}
\end{table}

\subsection{Dynamics are slowed when the angle is widened}
Widening the angle slows the dynamics of the liquid suggesting that a wide angle is preferred to make a good glass former. This is exemplified in Figure \ref{rotacf} showing the rotational correlation function (RCF)
\begin{equation} 
 C_\textrm{rot}(t)=\langle {\bf{u}}_i(t')\cdot{\bf{u}}_i(t'+t)\rangle
\end{equation} 
where ${\bf{u}}_i$ is a unit vector pointing along the longest bond of molecule $i$ and $\langle \ldots \rangle$ is average over molecules and $t'$. (This is a first order Legendre polynomial RCF and correspond to what is measured in dielectric spectroscopy \cite{hinze2004} if cross-correlations between molecules can be ignored). Figure \ref{tau} shows the temperature dependency of the characteristic rotational time $\tau_\textrm{rot}$ defined as
\begin{equation}
C_\textrm{rot}(\tau_\textrm{rot})\equiv 1/e.
\end{equation}

\begin{figure} 
\begin{center} 
  \includegraphics[width=1.0\columnwidth]{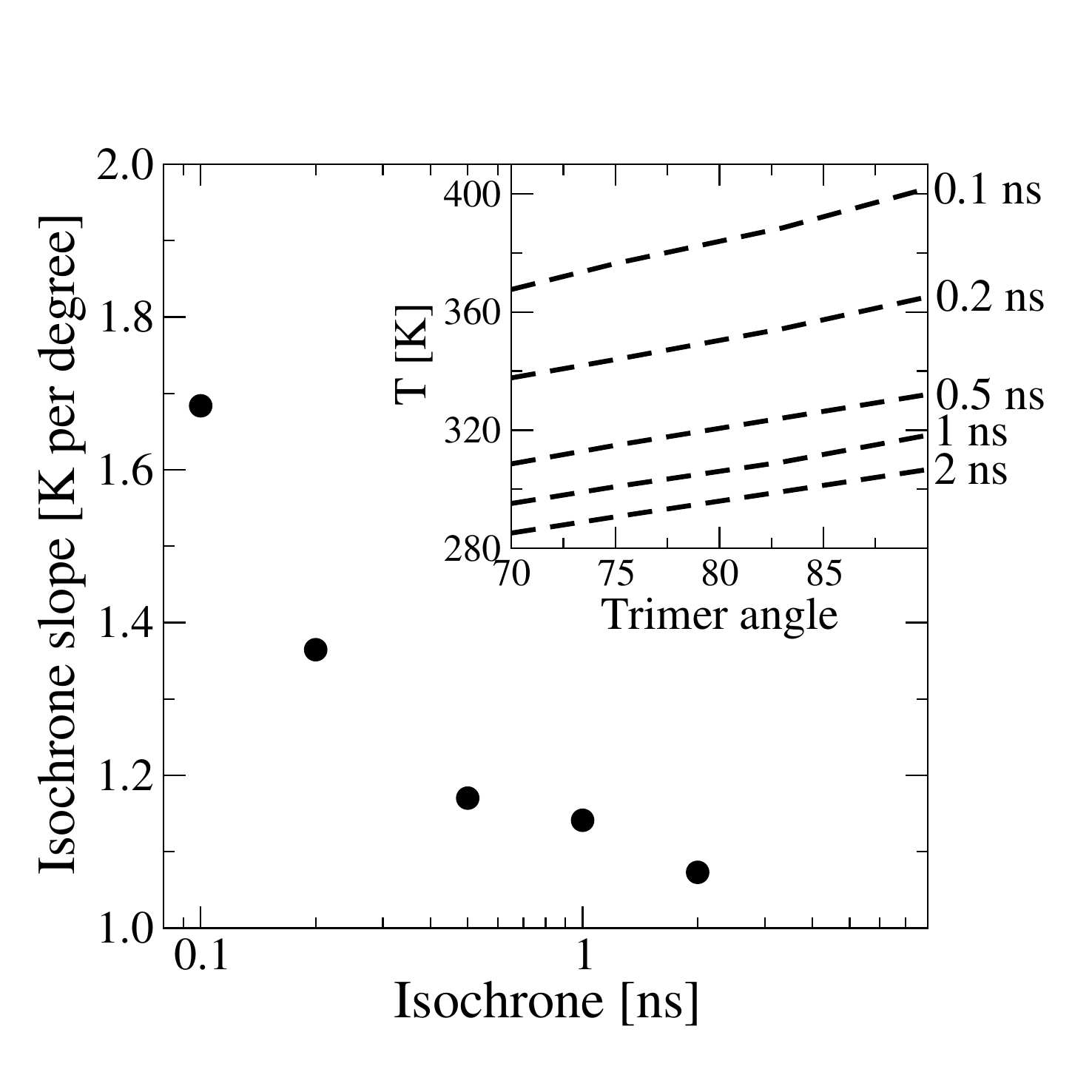}
\caption{\label{isochrone} The insert show rotational isochrones in the temperature-angle plane. Isochrones have a linear $\theta$ dependency with the slope shown in the main panel. At low temperature, the slope approaches 1 K per 1$^\circ$.}
\end{center} 
\end{figure}

How significant is the slowing down related to the widening of the angle relative to the lowering of the melting temperature?
To answer this, we investigate isochrones in the temperature-angle plane shown in the insert of Figure \ref{isochrone}. In the low-temperature region of the phase diagram, a angle increase of 1$^\circ$ correspond to a temperature increase of about 1 K. Assuming that this persists at lower temperatures, the glass-transition temperature $T_g$ have the same angle dependency.
For comparison, the change in melting temperature from 83$^\circ$ to 93$^\circ$ is 2.5 K per 1$^\circ$ (Figure \ref{trimerTm}). Thus, the change in melting temperature change more rapidly than isochrones (although, interestingly, changes are compatible in size for large angle changes).

\subsection{Predicting glass-forming ability from classical nucleation theory}
To relate these empirical findings to the glass-forming ability as a function of angle, we will in the following argue that the distance between the melting line and isochrones is an indicator of glass-forming ability.
Recall classical nucleation theory (CNT) where the prediction for the nucleation rate per unit volume is \cite{becker1935,gavezzotti2007}
\begin{equation} \label{CNT}
 k_\textrm{CNT} =\sqrt{\frac{\Delta \mu }{6\pi k_BTN_\textrm{c}}}\frac{24D_s N_\textrm{c}^\frac{2}{3}}{v_l \lambda^2}\exp\left\{-\frac{16\pi\gamma^3v_s}{3k_BT(\Delta\mu)^2}\right\}
\end{equation} 
where $D_s$ is the self-diffusion constant, $\lambda$ is a diffusion distance and
\begin{equation} 
N_\textrm{c}=-\frac{32}{3}\pi v_s\gamma^3(\Delta\mu)^{-3} 
\end{equation} 
is the size of the critical nucleus (in number of particles). In the super-cooled regime, $D_s$ varies exponentially or even super-exponentially with inverse temperature \cite{angell1985}. A convenient way to account for this dramatic change is to consider the rate along a isochrone $T_\tau(\theta)$ defined as a curve where the self-diffusion $D_s(T_\tau)\equiv1/\tau$ is constant (the glass-transition may be defined as such an isochrone). Along a $T_\tau(\theta)$ curve good glass-formers are characterized by having low nucleation rates. Near $T_m$ (where $\Delta\mu=0$ and the rate is zero) the dominant term in Equation (\ref{CNT}) is $\exp\{-(\Delta\mu)^{-2}\}$. Using that $\Delta\mu(T)\simeq[T_m-T]\Delta s(T_m)$ the nucleation rate per unit volume along the $\tau$-isochrone is
\begin{equation}\label{simpleCNT}
 k_\tau(\theta) = K_\tau(\theta) \exp\left\{-[T_m(\theta)-T_\tau(\theta)]^{-2}\right\}
\end{equation}
where $K_\tau(\theta)\simeq K_\tau$ is assumed near constant along isochrones. Thus, the angle that minimize $T_m(\theta)-T_\tau(\theta)$ may be used as a prediction of a good glass-former. Note that it is not crucial that CNT is quantitatively correct, but only qualitatively correct in the sense that $\Delta\mu$ changes dominates rate changes along isochrones. Equation (\ref{simpleCNT}) is convenient due to its simplicity, but approximations may be crude at temperatures where crystallization occurs. For a detailed investigation of this, crystallization rates could be computed using techniques for sampling rare events \cite{frenkel2002}. 

Using Equation (\ref{simpleCNT}), the empirical results suggest that the optimal glass-former angle is at about $\theta=83^\circ$ (determined graphically on Figure \ref{trimerTm} as the $\theta$ with the minimum distance between the melting line and the dashed red line). $T_\tau(\theta)$ have an significant angle dependency suggesting that better glass-formers may be found at angles larger than investigated in this study. Indeed extrapolation of the isochrone on Figure \ref{trimerTm} suggest that $\theta=100^\circ$ may be an good angle for a glass former (investigations of crystal structures that are stable at large angles would enlighten this hypothesis). Trimers with angles near $65^\circ$ (a little larger than the close-packing angle $60^\circ$) are expected to be prone to crystallization. This angle is optimal for the crystal, but not the liquid, suggesting that they may be more prone than the atomic LJ model. Slower liquid kinetics of trimers relative to the atomic liquid, however, may play an important 
role \cite{pedersen2011_lwotp}.

\section{Acknowledgments}
The author is grateful comments and suggestions from Christoph Dellago, Gerhard Kahl, Georg Kresse, Felix Hummel, Thomas B. Schr{\o}der, Jeppe C. Dyre and Peter Harrowell.

This work was supported by the VILLUM Foundation’s Matter (Grant No. 16515).


%
\end{thebibliography}

\end{document}